\documentclass[long, final]{jobim2013}

\usepackage[utf8]{inputenc}
\usepackage{multicol}
\usepackage{varioref}
\labelformat{figure}{\figurename~#1}
\labelformat{table}{\tablename~#1}

\labelformat{figure}{Fig.~#1}

\title{The genome of the medieval Black Death agent \\ extended abstract}

\titre{Le g\'enome de la bact\'erie responsable de la Peste Noire}

\author{Ashok Rajaraman\inst{1,2}, Eric Tannier\inst{3,4}, Cedric Chauve\inst{1,5}}

\institute{
  Department of Mathematics, Simon
  Fraser University, V5A 1S6 Burnaby (BC),
  Canada
 \\\email{\{arajaram,cedric.chauve\}@sfu.ca}
 \and
 International Graduate Training Center in
 Mathematical Biology, Pacific Institute for Mathematical Sciences,
 Vancouver (BC), Canada
 \and
 INRIA Grenoble
 Rh\^one-Alpes, F-38334 Montbonnot, France
 \\\email{eric.tannier@inria.fr}
 \and
 Universit\'e de Lyon 1, Laboratoire de Biom\'etrie et Biologie \'Evolutive,
 CNRS UMR5558 F-69622 Villeurbanne, France
 \and
 LaBRI, Universit\'e Bordeaux I, 33405 Talence, France
}

\abstract{%
  The genome of a 650 year old {\em Yersinia pestis} bacteria,
  responsible for the medieval Black Death, was recently sequenced and
  assembled into 2,105 contigs from the main chromosome. According to
  the point mutation record, the medieval bacteria could be an
  ancestor of most {\em Yersinia pestis} extant species, which opens
  the way to reconstructing the organization of these contigs using a
  comparative approach. We show that recent computational
  paleogenomics methods, aiming at reconstructing the organization of
  ancestral genomes from the comparison of extant genomes, can be used
  to correct, order and complete the contig set of the Black Death
  agent genome, providing a full chromosome sequence, at the
  nucleotide scale, of this ancient bacteria. This sequence suggests
  that a burst of mobile elements insertions predated the Black Death,
  leading to an exceptional genome plasticity and increase in
  rearrangement rate.
}

\keywords{Paleogenomics, computational biology, genome assembly, pathogens}

\resume{%
  R\'ecemment, le g\'enome d'une souche de la bact\'erie {\em Yersinia
    pestis} vieille de 650 ans a \'t\'e s\'equenc\'ee et assembl\'ee
  en 2,105 contigs issus de son chromosome. Cette bact\'erie
  m\'edi\'evale semble \^etre l'anc\^etre de la plupart des souches
  actuelles de {\em Yersinia pestis}, ce qui permet d'appliquer une
  approche comparative pour assembler ces contigs en scaffolds. En
  utilisant des m\'ethodes et principes r\'ecemment d\'evelopp\'es
  pour la reconstruction de l'organisation de g\'enomes anciens \`a
  partir de la comparaison de g\'enomes existants, nous corrigeons,
  organisons et compl\'etons les contigs de l'agent de la Peste Noire,
  pour obtenir une s\'equence compl\`ete pour le chromosome de cette
  bact\'erie ancienne. L'analyse de cette s\'equence sugg\`ere que de
  nombreuses insertions d'\'el\'ements mobiles ont particip\'e \`a
  l'\'emergence d'un g\'enome execeptionellement dynamique et \`a une
  augmentation du taux de r\'earrangements.
  }

\motscles{Pal\'eog\'enomique, bioinformatique, assemblage de g\'enomes, pathog\`enes.}

\begin{document}

\selectlanguage{english}

\begin{otherlanguage}{francais}
  \maketitle
\end{otherlanguage}

\section{Introduction}
\label{sec:introduction}

The plague has long been among the most feared human diseases
\cite{Camus1947}, due to dramatic pandemics such as the {\em Black
  Death} which ravaged Europe in the late middle-ages. Recently Bos
{\em et al.} \cite{Bos2011} were able to sequence the whole genome of
the Black Death agent, and concluded that it was an ancestor of most
extant strains of the human pathogen {\em Yersinia pestis} (see also
\cite{Schuenemann2011}).  The sequence extracted from the oral
metagenome of one individual was assembled using Velvet
\cite{Zerbino2008}, into approximately 130,000 contigs, including 2,105
contigs of length $\ge$ 500bp from the main chromosome, with similarities
with some {\em Yersinia} extant genomes\footnote{There are 2134 provided
contigs in total, and we discarded the 29 ones with no similarities with any {\em Yersinia}
extant genome because they are likely to be artefactual.}. This first
sequencing of the chromosome of an extinct prokaryote helped to
understand the causes of the Black Death pandemic
\cite{Bos2011,Parkhill2011,Wilson2012}.  However, the assembled 2,105
contigs cover only 85\% of the expected length of the ancestral
chromosome and their organization along this ancestral chromosome is
unknown, keeping out of reach a detailed genome-scale study of
the evolution of the structural organization of {\em Yersinia}
genomes, whose impact on pathogenicity is still an important open
question \cite{Chain2004}.

Current assembly methodologies can hardly be applied to fully assemble
and finish an ancient genome, aside of short molecules such as
plasmids \cite{Schuenemann2011} and organelle genomes
\cite{Paijmans2012}. Indeed, existing scaffolding methods, aimed at
ordering and orienting the contigs, and estimating the lengths of
inter-contig gaps, rely on additional data such as mate-pair reads
with mixed insert sizes
\cite{Bashir2012,Ribeiro2012,Gao2011,Salmela2011,Zerbino2009},
optical or physical maps \cite{Lin2012} or comparison with one or
several closely related genomes
\cite{Salzberg2008,Husemann2010}. However, due to the decay and
fragmentation of ancient DNA, reads from ancient genomes are in
general short, and optical maps or mate-pair libraries with long
inserts can not be obtained. This leaves the comparative approach as
the only possibility to scaffold large ancient genomes. The usual
setting of the comparative approach involves the comparison of the
contigs with one, or a few, closely related genomes, either genome
sequence or maps \cite{Blin2007,Bertrand2009,Munoz2010,Husemann2010} or
protein sequences \cite{Salzberg2008}. However, to the best of our
knowledge, none of these methods is intended to be applied on the
genome of an internal node of a given phylogeny.

\begin{figure}
\begin{center}
  \includegraphics[width=10cm]{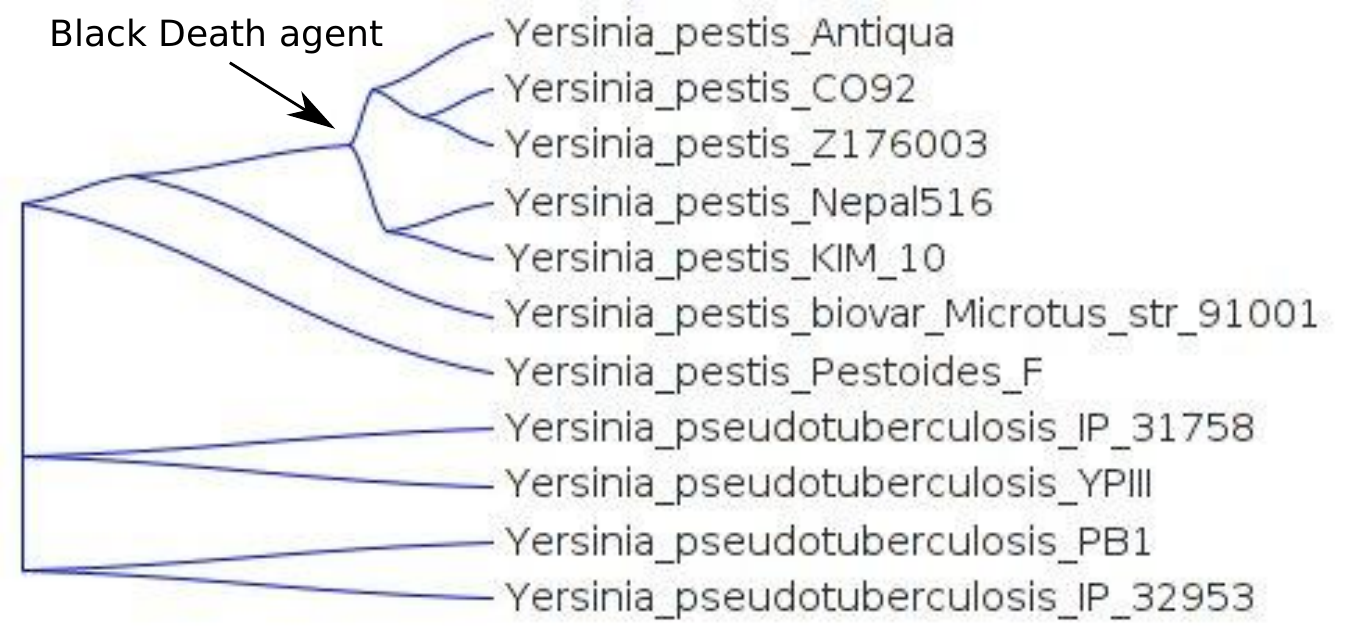}
  \caption{Phylogeny of the used extant genomes and position of the reconstructed one.}
  \label{fig:phylogenie}
  \end{center}
\end{figure}

We describe a comparative approach to scaffold an ancient genome, and
apply it to the medieval plague agent.  The ancestral Black Death
agent is indeed related to a dozen of descendants (from the {\em
  Yersinia pestis} clade) and close outgroups (from the {\em Yersinia
  pestis} and {\em Yersinia pseudotuberculosis} clades), whose
phylogeny, taken from Bos {\em et al.} \cite{Bos2011}, is shown on
\ref{fig:phylogenie}.

There has
been a recent flurry of ancestral genome organization reconstruction
methods, complementing classical methods for reconstructing ancestral
genome sequence \cite{Blanchette2004,Liberles2007,Diallo2010} and gene content
\cite{Cohen2010,Csuroes2010,Szollosi2012}. They have been used for
reconstructing ancestral genomes of bacterias
\cite{Wang2006,Fremez2007}, animals
\cite{Bourque2004,Ma2006,Putnam2007,Nakatani2007,Putnam2008,Chauve2008,Alekseyev2009,Muffato2010,Ouangraoua2011},
plants \cite{Sankoff2009,Murat2010}, yeasts
\cite{Gordon2009,Chauve2010,Bertrand2010} or protists
\cite{Ma2008}. Recent developments provide exact and fast algorithms
that handle sequence duplications, repeats, diverse types of genome
rearrangements and chromosome structures
\cite{Berard2012,Jones2012,Manuch2012}.

We show here that this corpus of methods is efficient and versatile
enough to be integrated into a comparative scaffolding framework for
ancient bacterial genomes, and we illustrate this claim with a
complete assembly of the medieval Black Death agent
chromosome. Starting from the contigs assembled by Bos {\em et al.}
\cite{Bos2011} which have similarities with extant {\em Yersinia}
genomes, we compute a single circular scaffold containing the ordered
and oriented sequences from the whole set of contigs, completed by
estimations of the sequences located between consecutive contigs
(gaps).  Additionally, we correct some contigs initially assembled by
Bos {\em et al.} by identifying probable {\em chimeric}, {\em
  redundant} or {\em duplicated} contigs. The chromosome structure we
observe is distant from every extant genome, explaining the difficulty
of the assembly process with a single reference genome. We annotate and analyse
the ancestral chromosome, pointing at a probable replication origin,
predicting the positions of insertion sequences (IS) and detecting the
numerous inversions that separate it from extant genomes.  We provide
evidence that the speciation between the {\em Yersinia pestis} and
{\em Yersinia pseudotuberculosis} clades was characterized by a burst
of insertion of IS elements in the {\em Yersinia pestis} genomes,
concomitant with an increase rate of genome rearrangements, which
breakpoints positions are also correlated with IS. 

\section{Results}\label{sec:results}

The main result of our work is a completely assembled chromosome
sequence of the Black Death agent genome. To obtain it, we followed a
generic procedure for reconstructing an ancestral genome organization
\cite{Ma2006,Chauve2008,Muffato2010,Bertrand2010,Jones2012}, which
comprises four phases: (1) extracting homologous families of ancestral and
extant genome markers, (2) computing putative linkage between ancestral
markers, (3) combining the set of ancestral linkages into a circular
sequence of ancestral markers, (4) infering inter-marker gap sequences.
We provide only a sketch of the implementation in this extended abstract,
and full details will be published elsewhere.

\paragraph{Families of homologous segments.} 
We aligned the ancestral contigs against 11 fully assembled genomes of
{\em Yersinia} strains. 
Several contigs were not aligned over their full length on every genome
because of rearrangements. So we cut the contigs into pieces, such
that every piece is aligned over its full length on every genome
and no pair of genomic segment defined by two different alignments
overlap (they are either disjoint or confounded).
This clusters ancestral and extant genome
segments into 2,619 homologous families. Each family
contains one or several ancestral contig segments, and zero, one or
several genome segments from each extant species.

All sequences from a single family are assumed to be homologous,
that is, they share a common ancestor and
having evolved through speciations, duplications, losses or
transfers. We do not have phylogenetic trees for the families that
would allow us to detect those events and derive a marker
content \cite{Szollosi2012}. Yet some ancestral markers correspond to
repeated sequences that were present at several loci of the ancestral
genome, while some of them contain ancestral segments from several
different contigs. We used phyletic profiles
\cite{Cohen2010,Csuroes2010} to determine the number of occurrences of
every ancestral marker, namely the ancestral marker content of this
ancestral genome.  We computed this ancestral content for each family
by using a parsimony approach that minimizes the number of gains and
losses of markers along the species tree for each family. This allows
to associate to each family a {\em multiplicity}, {\em i.e.} its
expected number of occurrences in the ancestral chromosome; 20
families out of 2,619 have a multiplicity greater than 1.

The amount of DNA encoded by the markers, when multiplicity is
 accounted for, is 3,846,866bp of ancestral DNA, while the initial
contigs encode 4,013,159bp. This initial loss of sequenced
 ancient DNA will be compensated by filling the gaps between the
 different pieces of the segmented contigs.

\paragraph{Computing putative linkages between ancestral markers.}
We computed sets of ancestral markers that are believed to be
consecutive in the ancestral chromosome. We call them {\em intervals}
of ancestral markers, if they contain more than two markers and {\em
  adjacencies} if they concern only two markers. We followed a
Dollo parsimony principle \cite{Chauve2008} to infer putative ancestral
linkages: a group of ancestral
markers is deemed to be contiguous in the ancestral genome
if markers from the same families are contiguous in at least two
extant genomes whose evolutionary path on the species phylogeny
contains the ancestor of
interest (here the Black Death agent). All 2,637 putative adjacencies
obtained in this way are then weighted according to their phylogenetic
conservation, using a recursive formula inspired from the
Fitch-Hartigan principle \cite{Ma2006,Chauve2008,Bertrand2010}.

\paragraph{Combining the set of ancestral linkages into a circular sequence
  of ancestral markers.} The set of putative ancestral adjacencies is
not compatible with a circular chromosomal structure, due to possible
converging genome rearrangements, for example. Indeed some markers
may be involved in too many adjacencies.
However, discarding 6
adjacencies out of the 2,637 putative ancestral adjacencies was enough
to obtain a set of maximal cumulative weight that can be ordered
circularly. They were found implementing a fast and exact "circularization"
method based on matching techniques in graphs\cite{Manuch2012}. 

\begin{figure}
  \begin{center}
    \includegraphics[width=0.5\columnwidth]{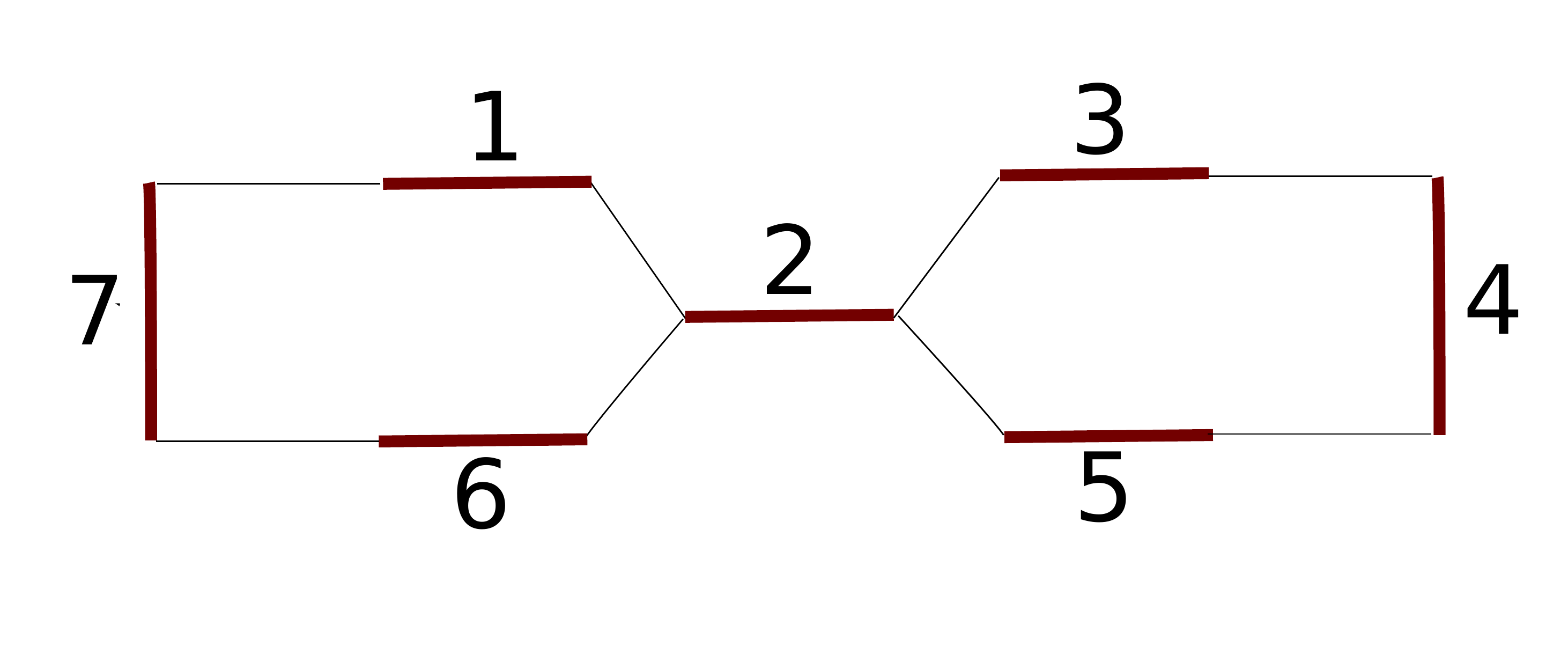}
    \caption{Illustration of the ambiguity in ordering ancestral markers
    with multiplicities greater than $1$ and of the use of intervals
    to address it.  Here is a toy example where we have markers
    $1,\dots,7$, drawn with bold red segments, and adjacencies between
    their extremities, drawn with thin black lines. Assume every
    marker has multiplicity $1$ except marker $2$, which has
    multiplicity 2. Then every marker extremity has as many
    adjacencies as its multiplicity predicts. But there are several
    possible circular orderings or these markers according to these
    adjacencies: 1,2,3,4,5,2,6,7, or 1,2,5,4,3,2,6,7. Suppose we have
    in addition size three intervals, and among them we find
    $\{1,2,3\}$ or $\{2,5,6\}$. Then only the first ordering is
    compatible. In our data set, intervals up to size $6$ were
    sufficient to completely clarify the adjacency signal.  }
  \label{fig:ambiguous-adjacencies}
\end{center}
\end{figure}

Adjacencies alone are compatible with many circular orders due to
repeated ancestral markers forming tangles in the adjacency graph
\cite{Husemann2010,Bashir2012}. To address this issue, intervals of
size greater than two were used as illustrated in
\ref{fig:ambiguous-adjacencies}
to clear the ambiguities, resulting in an ordering of the markers into
three large scaffolds.

We then joined the extremities of these three scaffolds to form a
circular chromosome by choosing, among the six possible
configurations, the only one supported by some extant genomes. This
resulted into a complete circular ordering of ancestral markers, where
each ancestral marker appears exactly as many times as it is expected
from its multiplicity.

\begin{figure}
  \includegraphics[width=\columnwidth]{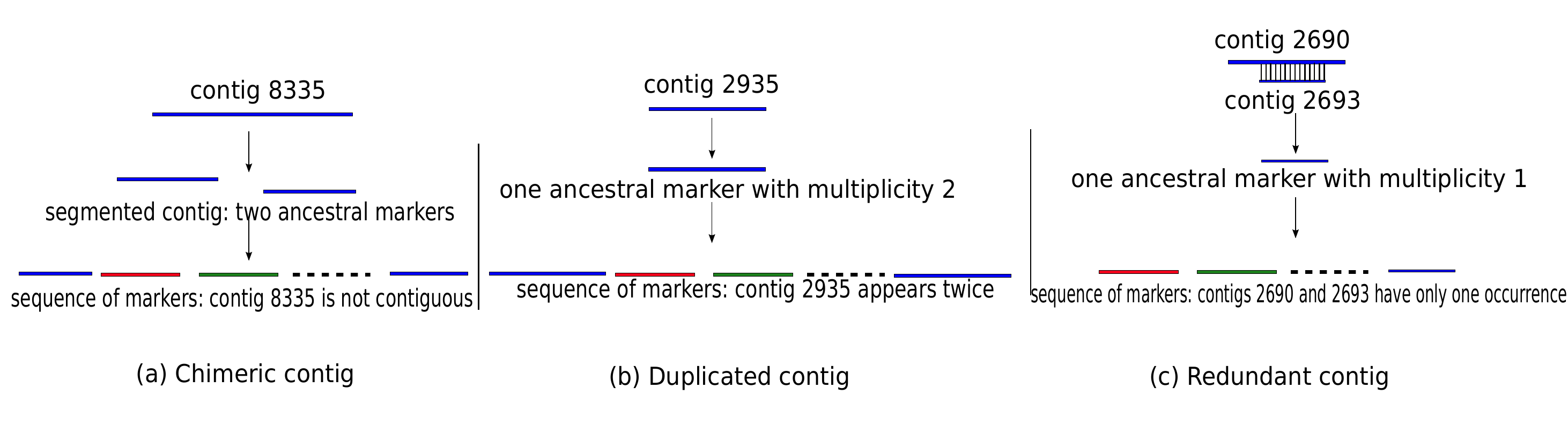}
  \caption{Contig correction: (a) the contig is cut during the
    segmentation procedure, but not joined during the marker ordering;
    (b) the contig is found to have two occurrences in the marker
    ordering; (c) two contigs contain the same DNA sequence and this
    sequence is predicted to have only one occurrence in the marker
    ordering.}
  \label{fig:correction}
\end{figure}

\paragraph{Correcting the initial contigs.}  In the resulting
ordering, each occurrence of an ancestral marker corresponds to one or
several segments of the initial contigs. The ordering of these
segments is mostly compatible with the initial contigs. We found only
one {\em chimeric} contig (see \ref{fig:correction}),
split into two non-adjacent markers in the ancestral genome
organization. None of the extant occurrences from the two families are
adjacent in extant genomes, pointing to either an assembly error
during the initial contig construction, or a derived rearrangement in
the ancient genome, which would be interesting since Bos {\em et al}
\cite{Bos2011} did not find such a mutation looking at nucleotide
substitutions. Note that the length filtering applied onto families
after the contig segmentation step can lead to an underestimation of
the number of chimeric contigs: if part of a contig has length less
than the threshold, it is discarded and the contig is not detected as
chimeric. Also four contigs segments were found to be {\em
  duplicated}: a large part ($>$ 500bp) of each is probably present in
more than one occurrence in the ancestral genome, while the initial
assembly predicted only one occurrence. Finally, 63 contigs have a
sequence which is found, up to very small variations, inside another
contig while their number of extant occurrences suggest they have
multiplicity one, so we believe they are {\em redundant}.

\paragraph{Estimating ancestral gaps sequences.}
We completed this assembly by estimating the sequences located in {\em
  ancestral gaps}, {\em i.e.} between pairs of ancestral markers
consecutive in the circular ordering.  For this we first estimated a
length interval for each ancestral gap: a length is said to be {\em
  supported} for an ancestral gap if there are two gaps in extant
genomes, in two species whose evolutionary path contains the ancestor
of interest, with such a length. The length interval of a gap is
defined by the minimum and maximum supported length for this gap. For
24 gaps we found no supported length, so we took the minimum and
maximum gap length of extant sequences in the species where the
markers are consecutive.  Then for each ancestral gap, we aligned all
extant gaps which lengths fall in the ancestral gap length
interval. We then constructed an ancestral sequence from each
alignment by an ancestral discrete character reconstruction method
implementing the Fitch algorithm \cite{Fitch1971}.

This resulted in an ancestral genome sequence of length 4,586,856
showing that 739,990bp were added to the ancestral markers sequences
by this finishing step. Only 1 gap was not assigned a sequence by this
method.


\paragraph{Analysis of the reconstructed ancestor.}
We took advantage of reconstructing
the full chromosome of the Black Death agent to analyze its structure and
evolution at the whole-genome scale.

\begin{figure}
  \includegraphics[width=\columnwidth]{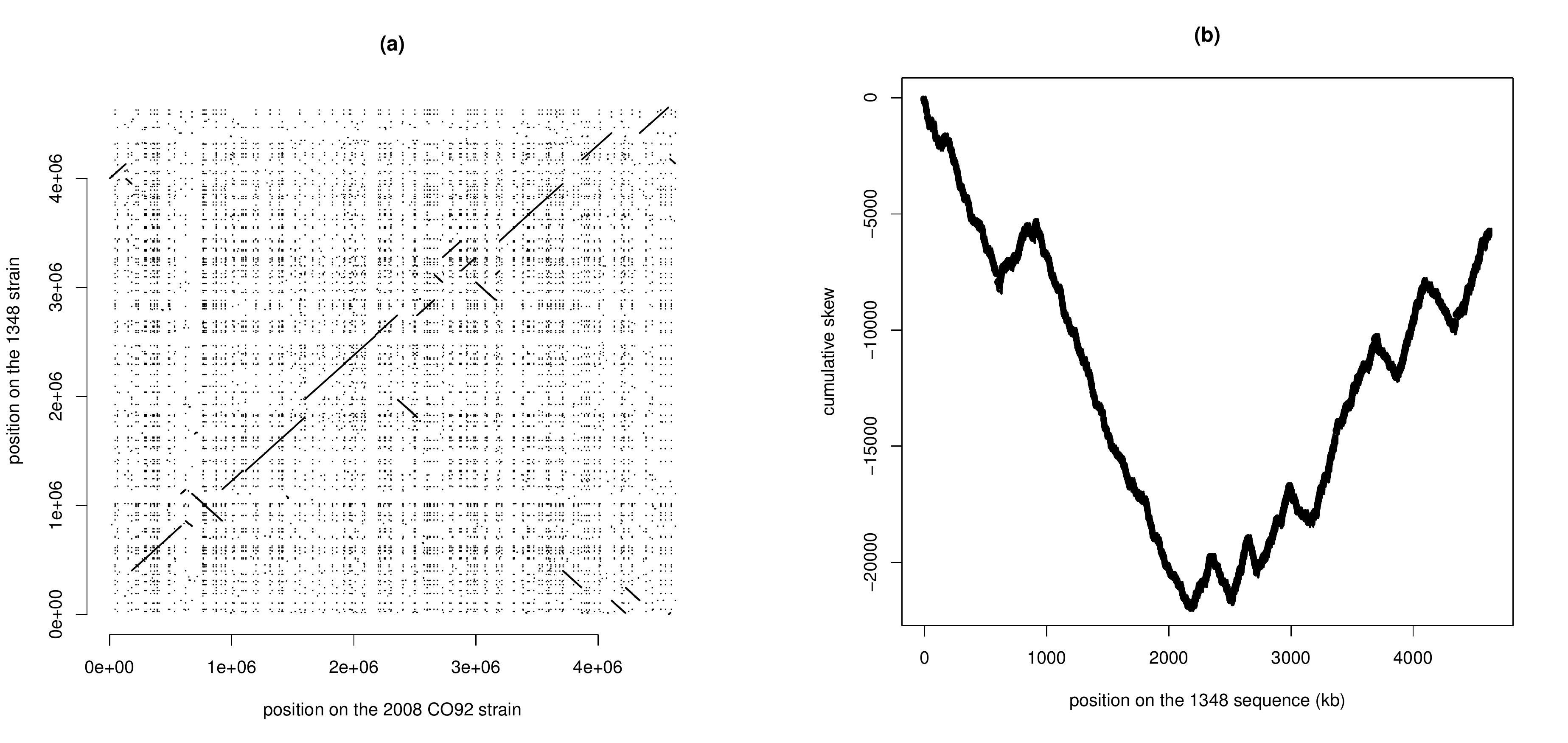}
  \caption{(a) Dotplot of all Megablast alignments of the medieval
    sequence against the CO92 extant strain.  The highly repetitive
    nature of both genomes appear, as well as the inversions that
    happened in the CO92 history, several of them being symmetric
    around the origin of replication. (b) Cumulative skew shows a
    probable position for the replication origin (for which we chose
    position 0), as well as the rearrangements which tend to blur the
    skew signal.}
  \label{fig:dotplot}
\end{figure}

We traced the GC-skew with SeqinR \cite{Charif2007} from a CDS
annotation by Glimmer (\ref{fig:dotplot}(b))
predict the position of the replication origin. We sliped the medieval
sequence such that the putative replication origin (the maximum value in
the cumulative GC-skew plot) has position 0 and we aligned the ancient
chromosome with the chromosome of the CO92 strain.  We obtained the
dotplot represented in \ref{fig:dotplot}(a)
that shows the highly repeated nature of both genomes, and the
rearrangements that have happened along the lineage from the ancestor
to the CO92 strain.

We mapped IS elements onto the reconstructed ancestral chromosome,
based on a conservative analysis of their patterns of presence in
extant markers and gaps: an ancestral gap is assigned an IS if one of
its occurrences in the descendants genomes is of length exactly the
minimum length of the ancestral gap and contains an annotated IS; we
focused on gaps as no extant marker does contain an annotated IS. This
resulted in 94 ancestral gaps containing IS. We confirmed this
comparative annotation with an automatic annotation.  Our analysis
also shows that a large part of these IS (at least 57) were already
present in the last common ancestor of all {\em Yersinia pestis}
strains, while they are almost completely absent from the genomes of
the considered {\em Yersinia pseudotuberculosis}.

We also analysed the genome rearrangements between the ancestral
sequence and extant genomes by sampling inversion scenarios between
the ancestral genome and the extant genomes (see \ref{fig:distances}). There are 8-9 inversions
between the {\em Yersinia pseudotuberculosis} strains and the medieval
genome, and 9-22 inversions when compared to (thought evolutionarily
closer) {\em Yersinia pestis} strains.  As noticed by Darling
et al \cite{Darling2008}, we can also observe that inversion
breakpoints are not randomly distributed and used: highly
used ones are concentrated in one third of the chromosome, around its
replication origin. Most inversions are symmetrical around the origin.
The positions of the inversion breakpoints are also highly correlated
with IS, as remarked earlier \cite{Deng2002}: 76 out of the 118 mapped
breakpoints are close ($<1000$bp distant) to some predicted IS, while
this number drops to 39 for uniformly sampled random coordinates
(p-value $<10^{-3}$). Rearrangements are very numerous in all {\em
  pestis} branches, strongly suggesting that they could be driven by
the IS.

\begin{figure}
\begin{center}
  \begin{tabular}{|l|c|}
    \hline
    Yersinia pestis biovar Microtus str 91001 &22\\
    Yersinia pestis Pestoides F &13\\
    Yersinia pseudotuberculosis IP 31758 &9\\
    Yersinia pseudotuberculosis YPIII &8\\
    Yersinia pseudotuberculosis PB1 &9\\
    Yersinia pseudotuberculosis IP 32953 &8\\
    Yersinia pestis Antiqua &21-22\\
    Yersinia pestis CO92 &12\\
    Yersinia pestis Z176003 &13\\
    Yersinia pestis Nepal516 &9\\
    Yersinia pestis KIM 10 &9\\
    \hline
  \end{tabular}
\end{center}
    \caption{Rearrangement distances between the extinct genome and the
    extant genomes. Two numbers mean that sampled scenarios have
    different length as we sample scenarios following a Bayesian
    posterior distribution of all scenarios, and not only the most
    parsimonious ones.}
  \label{fig:distances}

\end{figure}


\section{Discussion/Conclusion}
\label{sec:conclusion}

The present work illustrates the potential of phylogenetic/comparative
assembly methods to address the specific issues of ancient DNA
assembly (single reads, fragmentation, \dots). Our main result is a
complete assembly of the chromosopme of a 650 years old
bacteria, that opens the way to whole genome analysis of
rearrangements and insertion dynamics among others.

The method we developped for this assembly relies on recent advances,
both methodological and algorithmic, in reconstructing the
organization of ancient genomes from the comparison of related extant
genomes. We show here that such methods are generic enough to be also
used with data acquired by sequencing of ancient DNA. 

A crucial issue of such a method is its validation. In this extended abstract
we do not develop this point but we are currently extensively testing our
method on simulated data generated from {\em Yersinia} genomes.

We believe the methodological advances we present in this work
complement the recent breakthrough in ancient DNA sequencing, at least
for bacterial genomes, and suggest that integrating ancient genomes
into comparative genomics is an ambitious but realistic goal for the
next few years.

 \section*{Acknowledgements}
 \label{sec:acknowledgements}

 This work was supported by NSERC Discovery Grant to C.C., a PIMS IGTC
 Fellowship to A.R. and ANR-10-BINF-01-01 Ancestrome to E.T. We are
 thankful to Laurent Duret, Vincent Daubin, Annie Chateau, Eric
 Rivals, Hendrik Poinar for useful discussions.

\end{document}